\documentclass[aps,prl,twocolumn,showpacs]{revtex4}
\usepackage{amsmath,xcolor}
\usepackage{graphicx}

\newcommand{\ket}[1]{|#1\rangle}

\newcommand{\eq}{\begin{equation}}
\newcommand{\fine}{\end{equation}}

\begin{document}

\title{Complete and Deterministic discrimination of polarization Bell state assisted by momentum entanglement}
\author{M. Barbieri$^{1,\dag,*}$, G. Vallone$^{1,2,*}$, P. Mataloni$^{1,*}$ and \ F. De Martini$^{1,*}$ \\
$^{1}$Dipartimento di Fisica dell'Universit\'{a} ``La Sapienza'' and
Consorzio Nazionale Interuniversitario per le Scienze Fisiche della Materia,
Roma, 00185 Italy\\
$^2$Dipartimento di Fisica Teorica dell'Universit\`a di Torino and INFN -
sezione di Torino - 10100 Italy}

\begin{abstract}
A complete and deterministic Bell state measurement was realized by a simple
linear optics experimental scheme which adopts 2-photon
polarization-momentum hyperentanglement. The scheme, which is based on the
discrimination among the single photon Bell states of the hyperentangled
state, requires the adoption of standard single photon detectors. The four
polarization Bell states have been measured with average fidelity 
$F=0.889\pm0.010$ by using the linear momentum degree of freedom as the
ancilla. The feasibility of the scheme has been characterized as a function
of the purity of momentum entanglement.
\end{abstract}
\maketitle

In the domain of Quantum information (QI) the completion of most fundamental
quantum communication protocols involving bipartite entanglement, such as
quantum teleportation \cite{1}, quantum dense coding \cite{2}, entanglement
swapping \cite{3} and some important quantum cryptographic schemes \cite{4},
requires the complete and deterministic identification of the Bell states
which form the orthogonal basis for the reference Hilbert space of the
bipartite system.

In quantum optics, pairs of correlated photons are generated by spontaneous
parametric down conversion (SPDC) in a nonlinear (NL) optical crystal slab
by choosing suitably phase matching conditions. Photon qubits can be encoded
in several accessible degrees of freedom, such as polarization \cite{5,16},
linear and orbital momentum \cite{6,7}, and energy-time \cite{8,9}. 
In particular, the four orthogonal entangled Bell states,
expressed in the logic basis $\ket0,\ket1$:
\begin{equation}
\begin{aligned} \label{Bell} &|\Phi ^{\pm }\rangle = \frac{1}{\sqrt{2}}
\left(|0\rangle _{A}|0\rangle _{B}\pm |1\rangle _{A} |1\rangle _{B}\right),
\\ &|\Psi ^{\pm }\rangle =\frac{1}{\sqrt{2}}\left( |0\rangle _{A}|1\rangle_{B}
\pm |1\rangle _{A}|0\rangle _{B}\right)\,,
\end{aligned}
\end{equation}
form the complete maximally entangled basis of the Hilbert space
$\mathcal H_{A}\otimes \mathcal H_{B}$ with $dim(\mathcal H_{A})=dim(\mathcal H_{B})=2$. 
In the particular case of a
photon polarization entangled state, $\ket0$ and $\ket1$ correspond to the
horizontal ($\ket H$) and vertical ($\ket V$) polarization states.

By standard linear methods, the discrimination of polarization Bell states
can not be achieved by simply performing a single joint measurement on the
two particles. Indeed, a reliable experimental linear optical scheme capable to
deterministically distinguish among the four entangled Bell states with $100\%$ 
efficiency by using $2\times2$ entangled photon pairs, 
doesn't exist and only a partial Bell state analysis with a
maximum attainable value of $50\%$ efficiency can be performed \cite{10}.
Recently, probabilistic complete Bell state analyzers for photonic quantum
bits were demonstrated by using a controlled-Not (C-NOT) gate for photonic
qubits \cite{11}.

The strategy adopted to overcome the intrinsic probabilistic character of
any Bell analysis exploits further degrees of freedom to assist the 
measurement. In fact, by two photons entangled in $N>1$ degrees of freedom,
namely giving rise to an hyperentangled states 
spanning the $2^{N}\times 2^{N}$ Hilbert space, 
a complete and deterministic Bell state analysis can be performed
with standard linear optics \cite{12,13}. In the case of double entangled
states ($N=2$) it was shown that this operation can occur together with a
C-NOT logic operation between the control and target degrees of freedom \cite
{13}. An experimental demonstration of a complete analysis of the four
polarization entangled Bell states has been recently given by Schuck 
\textit{et al.} \cite{14}, who discriminate the polarization entanglement of two photons
generated by a Type II NL crystal assisted by the intrinsic time-energy entanglement
occurring in the SPDC process. The measurement apparatus described in that
work consisted of a sequence of three different steps which allowed to
distinguish among the four polarization entangled states. By that scheme, a
full deterministic analysis of all the photon pairs requires the adoption of
photon number resolving detectors.

In this letter we demonstrate that a complete and deterministic polarization ($\pi$) Bell state
analysis can be performed by using the further degree of freedom of momentum (\textbf{k}) as the ancilla. 
More precisely, the analysis of the Bell states (\ref{Bell}) is carried out
by discriminating among the single photon Bell states of a
$\pi$-$\mathbf{k}$ hyperentangled 2-photon
state, at the Alice ($A$) and Bob ($B$) sites. By our scheme the four Bell
states $|\Phi^{+}\rangle $, $|\Phi ^{-}\rangle $, $|\Psi ^{+}\rangle $, $%
|\Psi^{-}\rangle $, have been analyzed with high fidelity and equal
detection probabilities by a single step measurement apparatus and using
single photon detectors. On this purpose we used the SPDC source of $\pi$-$\mathbf{k}$
hyperentangled 2-photon states, based on a single Type I $%
\beta $-$BaB_{2}O_{4}$ (BBO) crystal, already described in other experiments
(cfr. Fig. 1a) \cite{15}.
\begin{figure}[t]
\includegraphics[scale=.43]{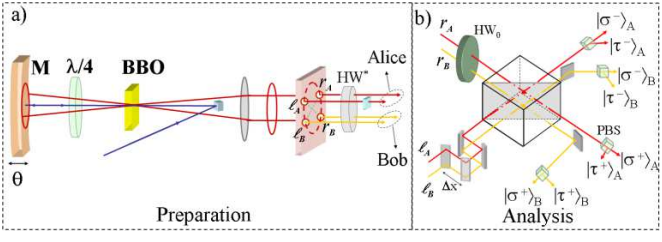}
\caption{a) Scheme of the hyperentanglement source: the polarization
entangled state $|\Phi \rangle = \frac{1}{\protect\sqrt{2}}\left( |H\rangle
|H\rangle +e^{i\protect\theta }|V\rangle |V\rangle \right) $ comes from the
superposition of the degenerate cones of a type-I BBO crystal. The basic
elements of the source are: [$i$] a spherical mirror $M$, reflecting both
the parametric radiation or the pump beam, whose micrometric displacement
allows to control the state phase $\protect\theta $ ($\protect\theta = 0,%
\protect\pi$). [$ii$] a $\protect\lambda /4$ waveplate, placed within the $M-
$BBO path, which performs the $|H\rangle_A|H\rangle_B \rightarrow
|V\rangle_A|V\rangle_B $ transformation on the 2-photon state belonging to
the left-cone. [$iii$] a positive lens which transforms the conical
parametric emission of the crystal into a cylindrical one. Mode selection is
performed by a four hole mask. The $\protect\lambda/2$ waveplate $HW^*$
intercepting modes $r_A,r_B$ performs the $|\Phi^\pm\rangle\rightarrow|\Psi^%
\pm\rangle$ transformation, the glass plate (on the $\ell_A$ mode) sets the
phase of the momentum state. b) Scheme of the Bell state analyzer (see text
for details). The delay $\Delta x$ is simultaneously varied for both $\ell_A$
and $\ell_B$ modes.}
\label{fig1}
\end{figure}
By this source we can generate over the whole BBO emission cone the
polarization entangled states. By inserting a four-holes screen aligned to intercept the
whole SPDC radiation, we select the photon pair passing through the modes
$\ell_A$-$r_B$ (left Alice-right Bob) or $r_A$-$\ell_B$,
with coherent superposition between the two events. 
Then the hyperentangled states
\begin{equation}
\begin{aligned} |\Xi \rangle &=|Bell\rangle_{AB} \otimes \left| \psi
^{+}\right\rangle\\ &=|Bell\rangle_{AB} \otimes \frac{1}{\sqrt{2}}\left(
|\ell\rangle_{A}|r\rangle_B+|r\rangle_A|\ell\rangle_B\right)
\label{hyper-ent} \end{aligned}
\end{equation}
can be generated \cite{15}. Here
the state $|Bell\rangle_{AB}$ can be either one of the 2-photon
polarization Bell states, $|\Phi ^{\pm }\rangle =\frac{1}{\sqrt{2}}\left(
|H\rangle _{A}|H\rangle _{B}\pm |V\rangle _{A}|V\rangle _{B}\right) $, $%
|\Psi ^{\pm }\rangle =\frac{1}{\sqrt{2}}\left( |H\rangle _{A}|V\rangle
_{B}\pm |V\rangle _{A}|H\rangle _{B}\right) $. 

The parametric source, which
allows to finely control the phase of the $\pi$-states, generates the hyperentangled
states $|\Phi ^{\pm }\rangle\otimes\ket{\psi^+} $ \cite{16}. The insertion of a
zero-order $\lambda /2$\ waveplate (wp) intercepting the modes $r_A$, $r_B$
(HW* in Fig. \ref{fig1}a) allows to transform the state $|\Phi ^{+}\rangle\otimes\ket{\psi^+}  $
in $|\Psi ^{+}\rangle\otimes\ket{\psi^+}  $, while the transformation $|\Phi ^{-}\rangle
\rightarrow |\Psi ^{-}\rangle $ is accompanied by a $\pi $ phase shift on
the momentum entangled state, $|\psi ^{+}\rangle \rightarrow |\psi
^{-}\rangle $. As a consequence, in order to generate $|\Psi ^{-}\rangle
\otimes |\psi ^{+}\rangle $, we need to compensate this phase shift by
suitable tilting of a thin glass plate inserted on mode $\ell_A$ (Fig. \ref
{fig1}a). The nonlocal character of the states $|\Xi \rangle $
was recently demonstrated by two different
experiments, the All Versus Nothing test \cite{17} and the Bell's
inequalities violation of local realism with two degrees of freedom \cite{18}.

By the present method, we are able to discriminate among the four possibility 
$|\Phi ^{+}\rangle $, $|\Phi ^{-}\rangle $, $|\Psi ^{+}\rangle$, 
$|\Psi ^{-}\rangle$, by using the single photon Bell basis:
\begin{equation}
\begin{aligned} 
\label{single-Bell} 
&\left| \sigma ^{\pm }\right\rangle _{i}
=\frac{1}{\sqrt{2}}\left[ |H\rangle |\ell\rangle _{i}\pm |V\rangle |r\rangle_{i}\right],\\
&\left| \tau ^{\pm }\right\rangle _{i}
=\frac{1}{\sqrt{2}}\left[ |V\rangle |\ell\rangle _{i}\pm |H\rangle |r\rangle_{i}\right] ,
\end{aligned}
\quad
i=A,B
\end{equation}
which allows to express the four possible states $|\Xi \rangle $ as
\begin{equation}
\begin{aligned}
|\Phi ^{\pm}\rangle \otimes |\psi ^{+}\rangle 
=\frac{1}{2}[&\pm|\sigma ^{+}\rangle _{A}|\tau ^{\pm}\rangle _{B}
\mp|\sigma ^{-}\rangle_{A}|\tau ^{\mp}\rangle _{B}+\\
&\quad+|\tau ^{+}\rangle _{A}|\sigma ^{\pm}\rangle_{B}
-|\tau^{-}\rangle _{A}|\sigma ^{\mp}\rangle _{B}],  \\
|\Psi ^{\pm}\rangle \otimes |\psi ^{+}\rangle  =\frac{1}{2}[&
\pm|\sigma ^{+}\rangle _{A}|\sigma ^{\pm}\rangle _{B}
\mp|\sigma ^{-}\rangle_{A}|\sigma ^{\mp}\rangle _{B}+\\
&\quad+|\tau ^{+}\rangle _{A}|\tau ^{\pm}\rangle_{B}
-|\tau^{-}\rangle _{A}|\tau ^{\mp}\rangle _{B}]\,.
\end{aligned}
\end{equation}
Each product state on the r.h.s. identifies unambiguously one of the states $\ket\Xi$.
Our scheme adopts linear momentum entanglement as the ancilla and
polarization entanglement as the target. It is equivalent to the one
proposed by Walborn et al. \cite{13}, except for the change of roles between
the momentum and polarization degrees of freedom in that case.
It is worth noting that by our scheme we distinguish among the four
hyperentangled states $|\Xi \rangle=|Bell\rangle_{AB} \otimes \ket{\psi^+}$.
However, since the momentum state $\ket{\psi^+}$ is fixed,
this is equivalent to distinguish among the four Bell polarization states.

Concerning the measurement apparatus, the two couples $\ell_A$-$r_B$ and $r_A
$-$\ell_B$ are spatially and temporally combined onto a $50\%$ beam splitter
($BS$) by an interferometric apparatus, where a trombone mirror assembly
with fine delay adjustment $\Delta x$ is mounted on the left ($\ell$) modes.
We set the position $\Delta x=0$ in correspondence of the
superposition between the mode pairs $\ell_A$-$r_B$ and
$r_A$-$\ell_B$, i.e. when the right ($r$) and left ($\ell$) optical paths of the
interferometer are equal \cite{15}. The analyzing apparatus is given by the $%
BS$ which follows a $45^{\circ }$ oriented $\lambda/2$ wp ($HW_{0}$),
inserted on the right ($r$) side in order to intercept both the Alice than
Bob modes (Fig. 1b) \cite{19}.

We are then able to completely distinguish among the states %
\eqref{single-Bell}, that are transformed by $HW_{0}$ as:
\begin{equation}
\begin{aligned} 
&\left| \sigma ^{\pm }\right\rangle _{i}\xrightarrow{\quad HW_0\quad}|H\rangle 
\otimes \frac{1}{\sqrt{2}}\left[ |\ell\rangle _{i}\pm|r\rangle _{i}\right],\\
&\left| \tau ^{\pm}\right\rangle _{i}\xrightarrow{\quad HW_0 \quad}|V\rangle \otimes
\frac{1}{\sqrt{2}}\left[ |\ell\rangle _{i}\pm |r\rangle _{i}\right]\,. 
\end{aligned}
\quad i=A,B
\end{equation}

The $BS$ discriminates between $|\ell\rangle _{A}+|r\rangle _{A}$ and $%
|\ell\rangle _{A}-|r\rangle _{A}$, $|\ell\rangle_{B}+|r\rangle _{B}$ and $%
|\ell\rangle _{B}-|r\rangle _{B}$ and polarization analysis on each $BS$
output mode, performed by a polarizing beamsplitter ($PBS$), completes the
single photon Bell state measurement \cite{19}. Note that a completely
deterministic Bell state analysis requires to detect the eight possible
outputs of the apparatus (Fig. 1b). In our proof of principle experiment we
used four avalanche single photon detectors (mod. Perkin Elmer SPCM-AQR14)
on the transmitted modes of the $PBS$'s. In the actual case the transmitted
polarization is set by a further $\lambda /2$ wp before each $PBS$.

We can also explain in a different way this effect: the hyperentangled states 
\eqref{hyper-ent} can be viewed as a three qubit states
\begin{equation}\label{3qubit}
|\Xi \rangle =|Bell\rangle _{AB}\otimes \frac{1}{\sqrt{2}}\left( |0\rangle
_{C}+|1\rangle _{C}\right)
\end{equation}
where now the qubit $C$ is represented by the \textit{couple of photons} in
the coherent superposition of the two states $|0\rangle _{C}=|\ell \rangle
_{A}|r\rangle _{B}$ and $|1\rangle _{C}=|r\rangle _{A}|\ell \rangle _{B}$ .
We are then able to completely discriminate
between the four polarization Bell states $|Bell\rangle _{AB}$ of the two
qubits $A$ and $B$ with the a priori information about the state of the
ancillary qubit $C$. This is the minimum a priori information (one over
three qubits) required to perform a complete and deterministic Bell state
analysis by linear optics. It is well known that this discrimination is not
possible with only two qubits and no extra information \cite{10}.
Our approach improves the ``standard'' Bell state analysis where two bits of
information are contained in the four Bell states and just one bit,
concerning the information on {which of the two kinds of states, }$|\Phi
\rangle ^{\pm }${\ or }$|\Psi \rangle ^{\pm }$, the input particles are in,
can be deterministically and completely extracted.
It is worth noting the relevance
for communication or cryptographic protocols of our method which allows to extract all 
(i.e. two) the bits of information that can be encoded in the states \eqref{3qubit}
\footnote{The present method is feasible only for 
quantum communication schemes, e.g. dense coding, which do not require independent photons.}.

Hence, Bell state analysis is performed by the following procedure:

\begin{itemize}
\item[1)]  The phase information (the $+$ or $-$ signs) of the Bell states
is transferred  into the qubit $C$. In fact the $HW_{0}$
operates in the following way:
\begin{equation}
\begin{aligned}
&|\Phi^\pm\rangle_{AB}\otimes\ket{+}_C\xrightarrow{\quad HW_0\quad}
|\Psi^\pm\rangle_{AB}\otimes\ket{\pm}_C\\
&|\Psi^\pm\rangle_{AB}\otimes\ket{+}_C\xrightarrow{\quad HW_0\quad}
|\Phi^\pm\rangle_{AB}\otimes\ket{\pm}_C\end{aligned}
\end{equation}
where $\ket{\pm}_C=\frac1{\sqrt2}\left(|0\rangle_C\pm|1\rangle_C\right)$.

\item[2)]  The $BS$ discriminates between $\ket{+}_C$ and $\ket-_C$ as follows:
the photons emerge either on the same or the opposite sides of the $BS$
depending of the states $\ket+_C$ or $\ket-_C$, respectively.

\item[3)] The four PBSs perform polarization analysis distinguishing between $|\Psi
\rangle $ and $|\Phi \rangle $.
\end{itemize}

\begin{figure}[t]
\includegraphics[scale=.9]{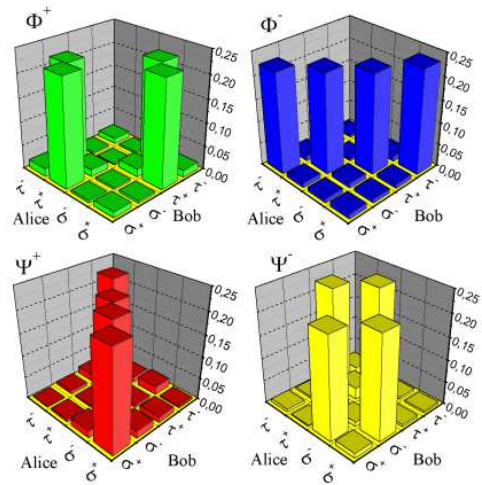}
\caption{Experimental coincidence frequencies showing the complete Bell
state analysis of the polarization states $|\Phi ^{+}\rangle $, $|\Phi
^{-}\rangle $, $|\Psi ^{+}\rangle $, $|\Psi ^{-}\rangle $. Relative errors
are typically 2\% for the maxima, 5\% for the other terms.}
\label{fig2}
\end{figure}
The four 3-D histograms given in Fig. \ref{fig2} show all the $16$ possible
combinations of the states (\ref{single-Bell}) for either one of the input
states $|\Phi ^{+}\rangle $, $|\Phi ^{-}\rangle $, $|\Psi ^{+}\rangle $, $%
|\Psi ^{-}\rangle $ and demonstrate the successful implementation of the
Bell state analyzer. Each datum was obtained in an acquisition time of $%
10\sec $, while the typical count rate was $\simeq 1000\sec ^{-1}$ for each
maximum measurement.
\begin{figure}[t]
\includegraphics[scale=.8]{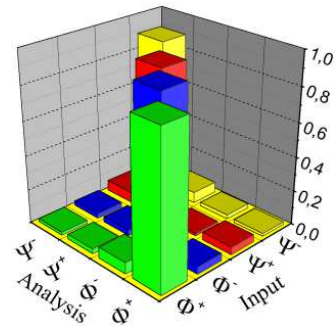}
\caption{Overall experimental fidelities obtained by the Bell state analyzer
for each input Bell state. Relative errors are typically 2\% for the maxima,
5\% for the other terms.}
\label{fig3}
\end{figure}
The overall input-output histogram shown in Fig. \ref{fig3} clearly
indicates the high efficiency of the analysis performed by our scheme. The
achieved fidelities of each Bell-state analysis are $F_{\left| \Phi
^{+}\right\rangle }=0.886\pm 0.018$, $F_{|\Phi ^{-}\rangle }=0.895\pm 0.018$%
, $F_{|\Psi ^{+}\rangle }=0.877\pm 0.018$, $F_{|\Psi ^{-}\rangle }=0.899\pm
0.018$, with an average value of $0.889\pm 0.010$. Note that
the adoption of the same measurement apparatus allows to identify the four
Bell states with almost the same fidelity. The noise contribution due to the
unexpected coincidences is partially caused by the non perfect purity of the
polarization input state and partially due to imperfections of the analysis
set-up, e.g. mode mismatch on $BS$.
\begin{figure}[t]
\includegraphics[scale=.8]{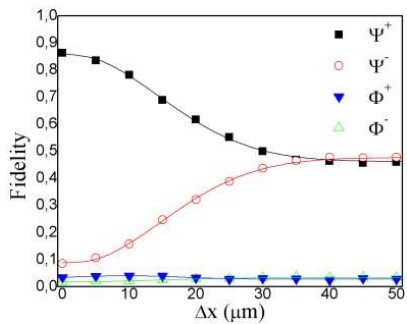}
\caption{Output fidelities of the states $|\Phi ^{+}\rangle $, $|\Phi
^{-}\rangle $, $|\Psi ^{+}\rangle $, $|\Psi ^{-}\rangle $, vs. the path
length difference $\Delta x$ in the interferometric apparatus (Input state: $%
|\Psi ^{+}\rangle $). Error bars are smaller than the corresponding
experimental points.}
\label{fig4}
\end{figure}

To test the feasibility of the Bell state analyzer realized by our scheme,
we measured the output of the analyzer when the state $|\Psi ^{+}\rangle $
is injected, while introducing noise in a controlled way in the ancilla
state $|\psi ^{+}\rangle $. This was performed by varying the value of $%
\Delta x$ in the interferometric apparatus. This procedure 
makes the two events, corresponding to the photons passing through the modes 
$\ell _{A}$-$r_{B}$ or $r_{A}$-$\ell _{B}$, more
distinguishable and simulates an increasing amount of decoherence 
between the two possible mode pairs (not between one photon and the other). 
As a consequence the final state is pure in
polarization and mixed in the momentum degree of freedom. The experimental
output fidelities, shown in Fig. \ref{fig4}, indicate, as expected, that $%
|\Psi ^{+}\rangle $ and $|\Psi ^{-}\rangle $ can not be discriminated when $%
\Delta x>l_{coh}$, the coherence length of the down converted photons
imposed by the $\Delta \lambda =6$nm interference filters before the
detectors. The results of Fig. 4 demonstrate that a still efficient Bell
state analysis, with $F_{\left| \Psi ^{+}\right\rangle }\geq 0.75$, may be
performed even with a partially degraded ancilla state. Similar results are
expected when the input polarization entangled state is partially mixed.

We have presented in this letter a linear optical scheme based on two photon
hyperentanglement which allows to perform in a deterministic way the
simultaneous measurement of the four polarization Bell states by using
standard single photon detectors. By virtue of the simplicity of the
measurement procedure and of the high fidelity experimentally attained, the
present Bell state analyzer (Fig.1b) may be applied to any source able to
produce polarization-momentum entangled photons \cite{5,15,20} and could be
useful for the realization of QI protocols, in particular dense coding and
quantum key distribution.
Precisely, the implementation of cryptographic schemes with qudits up to $%
d=4$ (ququarts) requiring $5$ mutually unbiased bases and the consequent
Bell state measurement can be efficiently performed by adopting the method
described in the present work \cite{21}. Indeed, it has been shown that
these systems are more robust against specific classes of eavesdropping
attacks \cite{pasquin}. 

Thanks are due to Serge Massar for useful discussions. This work was
supported by the FIRB 2001 (\textit{Realization of Quantum Teleportation and
Quantum Cloning by the Optical Parametric Squeezing Process}) and PRIN 2005 (%
\textit{New perspectives in entanglement and hyper-entanglement generation
and manipulation}) of MIUR (Italy).
\\\\
*Web-page: http://quantumoptics.phys.uniroma1.it/\\
$\ ^\dag$present address: School of Physical Sciences, the University of
Queensland, 4072, Brisbane, QLD, Australia.


\begin{thebibliography}{99}
\bibitem{1}  C. H. Bennett \textit{et al.}, \textit{Phys. Rev. Lett.}
\textbf{70}, 1895 (1993).

\bibitem{2}  C. H. Bennett and S. J. Wiesner, \textit{Phys. Rev. Lett.}
\textbf{69}, 2881 (1992).

\bibitem{3}  T. Jennewein, G. Weihs, J. W. Pan, and A. Zeilinger, \textit{%
Phys. Rev. Lett.} \textbf{88}, 017903 (2002); F. Sciarrino, E. Lombardi, G.
Milani, and F. De Martini, \textit{Phys. Rev. A} \textbf{66}, 024309 (2002).

\bibitem{4}  A.Ekert, \textit{Nature} \textbf{358}, 14 (1992); N. Gisin, G.
Ribordy, W. Tittel, and H. Zbinden, \textit{Rev. Mod. Phys.} \textbf{74},
145 (2002).

\bibitem{5}  P. G. Kwiat \textit{et al.}, \textit{Phys. Rev. Lett.} \textbf{%
75}, 4337 (1995); P. G. Kwiat, E. Waks, A. G. White, I. Appelbaum and P. H.
Eberhard, \textit{Phys. Rev. A} \textbf{60}, R773 (1999).

\bibitem{16}  C. Cinelli, G. Di Nepi, F. De Martini, M. Barbieri, and P.
Mataloni, \textit{Phys. Rev. A} \textbf{70}, 022321 (2004).

\bibitem{6}  J.G.Rarity, P.R.Tapster, \textit{Phys. Rev. Lett.} \textbf{64},
2495 (1990).

\bibitem{7}  N. Langford \textit{et al.}, \textit{Phys. Rev. Lett.} \textbf{%
93}, 053601 (2004).

\bibitem{8}  J. D. Franson, \textit{Phys. Rev. Lett.} \textbf{62}, 2205
(1989).

\bibitem{9}  J. Brendel\textit{\ et al.}, \textit{Phys. Rev. Lett.} \textbf{%
82}, 2594 (1999).

\bibitem{10}  N. L\"utkenhaus, J. Calsamiglia, and K.-A. Suominen, \textit{%
Phys. Rev. A} \textbf{59}, 3295 (1999); J. Calsamiglia and N. L\"utkenhaus,
\textit{Appl. Phys. B} \textbf{72}, 67 (2001).

\bibitem{11}  J. L. O'Brien \textit{et al.}, \textit{Phys. Rev. Lett}
\textbf{93}, 080502 (2004); Z. Zhao \textit{et al.}, \textit{Phys. Rev. Lett}
\textbf{94},030501 (2005); P. Walther and A. Zeilinger, \textit{Phys. Rev. A}
\textbf{72}, 010302(R) (2005).

\bibitem{12}  P. G. Kwiat and H. Weinfurter, \textit{Phys. Rev. A},\textbf{\
58}, R2623 (1998).

\bibitem{13}  S. P. Walborn, S. P\`{a}dua and C. H. Monken, \textit{Phys.
Rev. A }\textbf{68}, 042313 (2003).
	
\bibitem{14}  C. Schuck, G. Huber, C. Kurtsiefer, and H. Weinfurter, \textit{%
Phys. Rev. Lett.} \textbf{96}, 190501 (2006).

\bibitem{15}  M. Barbieri \textit{et al.}, \textit{Phys. Rev. A} \textbf{72}%
, 052110 (2005).

\bibitem{17}  C. Cinelli \textit{et al., Phys Rev. Lett.} \textbf{95},
240405 (2005).

\bibitem{18}  M. Barbieri \textit{et al.}, 
\textit{Phys. Rev. Lett.} {\bf 97}, 140407 (2006)

\bibitem{19}  D. Boschi, S. Branca, F. De Martini, L. Hardy, S. Popescu,
\textit{Phys. Rev. Lett.} \textbf{80}, 1121 (1998).


\bibitem{20}  T.Yang \textit{at al. , Phys. Rev. Lett.} \textbf{95}, 240406
(2005).

\bibitem{21}  G. M. D'Ariano, P. Mataloni and M. F. Sacchi \textit{Phys. Rev.
A } \textbf{71}, 062337 (2005).

\bibitem{pasquin} H. Bechmann-Pasquinucci and A. Peres {\it Phys. Rev. Lett.} {\bf 85}, 3313 (2000) 

\end{thebibliography}
\end{document}